\documentclass[twocolumn,aps,prl,showpacs,floatfix]{revtex4}
\usepackage{amsmath}
\usepackage{amsfonts}
\usepackage{amssymb}
\usepackage{bm}
\usepackage{graphicx}
\begin{document}

\title{Species clustering in competitive Lotka-Volterra models}

\author{Simone Pigolotti}
\author{Crist\'obal L\'opez}
\author{Emilio Hern\'andez-Garc\'\i a}
\affiliation{Instituto de F{\'\i}sica Interdisciplinar y Sistemas
Complejos IFISC (CSIC-UIB), Campus de la Universitat de les Illes
Balears, E-07122 Palma de Mallorca, Spain. }

\date{\today}

\begin{abstract}

We study the properties of Lotka-Volterra competitive models in
which the intensity of the interaction among species depends on
their position along an abstract niche space through a competition
kernel. We show analytically and numerically that the properties
of these models change dramatically when the Fourier transform of
this kernel is not positive definite, due to a pattern forming
instability. We estimate properties of the species distributions,
such as the steady number of species and their spacings, for
different types of kernels.

\end{abstract}


\maketitle

It is widely believed that competition among species greatly
influences global features of ecosystems. One of the most relevant
is the fact that ecosystems can host a limited number of species.
The common explanation is the so-called limiting similarity
\cite{macarthurlevins} and involves representing species as points
in an abstract {\it niche space}, whose coordinates quantify the
phenotypic traits of a species which are relevant for the
consumption of resources, usually the typical size of individuals,
but also preferred prey, optimal temperature and so on. On general
grounds, one expects that a species experiences a stronger
competition with the closer species in this space. As a
consequence, a species can survive if it is able to maintain its
distance with the others above a minimum value which depends on
the intensity of competition. On the contrary, when the distance
between two species becomes too small, the unavoidable difference
in how efficiently the two species feed on the resources will make
one of the two outcompete the other, that will go extinct: this is
the essence of the competitive exclusion principle \cite{hardin},
and it can be thought as the theoretical grounding of the concept
of ecological niche. Thus, one expects a stable ecosystem to
display a finite number of species, more or less equidistant in
the niche space. The finiteness of the number of species has been
observed in several competition models \cite{szabo}  and,
recently, it has been rigorously demonstrated for a general class
of them \cite{gyllenberg}.

Deviations from the above scenario have aroused renewed interest
recently, when it was observed numerically \cite{schefferpnas}
that the equilibrium state of rather standard models is not always
characterized by a homogeneous distribution of species in niche
space. Instead, clumpy distributions, with clusters of many
species separated by unoccupied regions, are observed. Evidences
of a similar phenomenon have been observed recently in an
evolutionary model \cite{brigatti}, suggesting that a theoretical
explanation of these patterns could bring new insights in the
study of speciation mechanisms \cite{johansson}.

In this Letter, we study the Lotka-Volterra (LV) competitive model
as the prototype of competitive systems. While the statistical
properties of the antisymmetric LV model have been addressed
recently \cite{tokitaprl}, not much has been demonstrated about
the statistics of the competitive case. Through this paper we will
use the language of ecological species competition, but we stress
that the LV set of equations appears in contexts as diverse as
multimode dynamics in optical systems \cite{optics}, technology
substitution \cite{tech}, or mode interaction in crystallization
fronts \cite{Kurtze}, and it is the natural starting point when
modelling competitive systems. Our main result is that the
macroscopic clustering of species is related to a pattern forming
transition that separates two different qualitative behaviors.
This is the same phenomenology found in birth-death particle
systems with interaction at a distance, in which individuals
typically aggregate forming clusters which arrange in an ordered
pattern \cite{refbugs1}, with the physical space playing the role
of niche space. The feature which is relevant
for this transition is not the intensity of the competition, but
the functional form of the competition kernel. We estimate
analytically the number of species in cases at both sides of the
transition, and also discuss the role of species heterogeneity.

Mathematically, one has competition when the growth of a species
affects negatively the growth rate of other species. One of the
simplest implementations of that is the LV competitive model:
\begin{equation}\label{LVmodel}
\dot{n_i}=n_i\left(r-a_i\sum_{j=1}^N g(|x_i-x_j|)n_j \right)\
,\qquad  \ i=1\ldots N.
\end{equation}
$N$ is the number of species and $n_i$ denotes the population of
species $i$. Each species is characterized by a growth rate $r$
and a competition parameter $a_i$ (we take into account
differences among species only in the latter parameter). A species
is also characterized by a position $x_i$ in a niche space that we
assume, for simplicity, to be the segment $[0,L]$ with periodic
boundary conditions. Generalization to a multidimensional niche
space is straightforward, and we expect the unrealistic boundary
conditions assumed here to be irrelevant except close to the
interval endpoints. The competition kernel $g(x)$ is a
non-negative and non-increasing function. Note that the sum in
Eq.(\ref{LVmodel}) contains the self-interaction term $g(0)n_i$.

To fully specify the dynamics, we should state how the $x_i$ are assigned to
species and eventually changed.  We consider a slow immigration rate $I$ at
which new species, characterized by a random phenotype $x \in [0,L]$, are
introduced in the system with a small random population $\delta n$. This
choice is appropriate to model a situation like an ecological community on
an island \cite{macarthurwilson}. In addition, we consider extinct, and
remove from the system, species whose population goes below a given
threshold $n_T$. When $I^{-1}$ is very large compared to the timescales of
population dynamics, the system has time to relax to a quasisteady state
after each immigration event. Our interest here is in the characteristics
of these states, in which immigration plays almost no role. An efficient
way to obtain them is by running simulations of Eq. (\ref{LVmodel}) with a
small amount of immigration, which is then ``switched off'' after some
time. By a choice of the time and the population units, we can set $r=1$ and
$n_T=1$ (thus the parameters $a_i$ are really $a_i n_T/r$). We also measure
all distances in niche space in units of $L$, so that $L=1$. 
\begin{figure}[tbh]
\includegraphics[width=\columnwidth,clip=true]{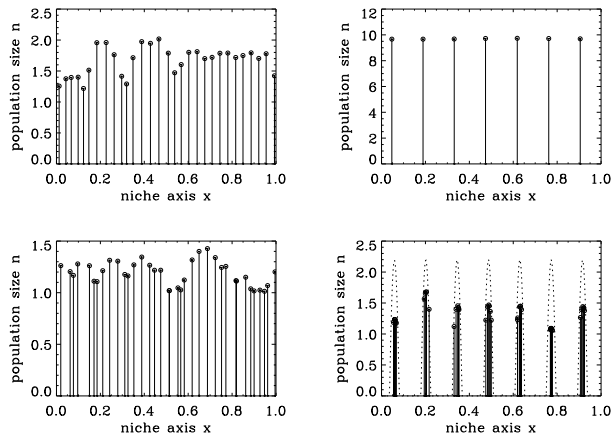}
\caption{\label{fig:patterns} Steady states of system
(\ref{LVmodel}), with $a_i=a=0.1$, obtained by evolving a random
configuration, initially of $200$ species, for a time $t=5\times
10^5$. The immigration rate was initially $I=0.004$, and switched
off after half simulation. Top panels: Competition kernel
$g_1(x)=\exp(-x/R)$ (left), and $g_4(x)=\exp[-(x/R)^4]$ (right),
with $R=0.1$. In the bottom panels, we added a Kronecker delta
$\delta_{x0}$ to the kernels above. In the last panel, the
dotted line is a steady solution of (\ref{fieldtheory}),
arbitrarily scaled in the vertical to fit in the same plot.}
\end{figure}

In Fig.~\ref{fig:patterns} we show numerical results for the
dynamics just defined, for $a_i=a$ for all $i$. In the top row we
compare the distribution of species with kernels
$g_1(x)=\exp(-x/R)$ and $g_4(x)=\exp[-(x/R)^4]$. In the
exponential case (left) species occupy the full niche space.
Although they are not perfectly equispaced and there are
differences in the population sizes, there is a clear average
interspecies distance, which corresponds to $1/N$. In the
quartic-exponential case (right), a much more regular pattern
emerges, with different species perfectly equidistant (and all
with the same population). Since growth limitation is known to
affect species distribution \cite{schefferpnas}, we plot in the
bottom row results for kernels $g_1(x)+\delta_{x0}$ and
$g_4(x)+\delta_{x0}$, i.e. the same kernels but with an enhanced
value of the self-competition coefficient $g(0)$. Here the
difference is even more striking: the exponential case is similar
to the previous one, but the quartic case shows clear clusters of
species separated by empty regions.

To understand the origin of the periodic patterns, we write a
continuum evolution equation for the field $\phi(x,t)$, the
expected density of individuals in a given point $x$ of the niche
space as a function of time:
\begin{equation}\label{fieldtheory}
\partial_t \phi(x,t)=\phi(x,t)\left(1-a\int g(|x-y|)\phi(y,t)dy \right)+s,
\end{equation}
which is a mean field version of Eq.~(\ref{LVmodel}) for $a_i=a$.  In this
{\it macroscopic} description, we neglect fluctuations in the immigration
process by using a constant rate $s=I\delta n$.  The stationary homogeneous
solutions of Eq.~(\ref{fieldtheory}) are
$\phi_0=(1\pm\sqrt{1+4s\hat{a}})/(2\hat{a})$, where $\hat{a}=a\mathcal{N}$,
with $\mathcal{N}=\int g(x)dx$. Of the two solutions, only the one with the
plus sign is acceptable since the other leads to a negative density (this
second solution corresponds to the extinct absorbing state when $s=0$). We
analyze the stability of the positive solution by considering a small harmonic
perturbation $\phi=\phi_0+\epsilon\exp(\lambda t+ikx)$.  Substituting into
(\ref{fieldtheory}), the first order in $\epsilon$ gives the following
dispersion relation:
\begin{equation}\label{dispersion}
\lambda(k)=1-\phi_0\hat a \left(1+\frac{\tilde{g}(k)}{\mathcal{N}}
\right)
\end{equation}
where $\tilde{g}(k)=\int g(x)\exp(-ikx)dx$ is the Fourier
transform of $g(x)$. When $\lambda$ becomes positive for some
values of $k$, the constant solution of (\ref{fieldtheory}) is
unstable, signaling a pattern forming transition \cite{CrossHoh}
with the characteristic length scale of the pattern determined by
the value of $k$ at which $\lambda(k)$ is maximum. For
Eq.~(\ref{dispersion}), in the limit $s\to 0$, it is a sufficient
and necessary condition for instability that the Fourier transform
of the kernel, $\tilde g$, takes negative values (notice that
$\phi_0\hat a\geq 1$, with $\phi_0\hat{a}\sim1$ in the limit
$s\rightarrow 0$; for sufficiently large $s$, the homogeneous
state regains stability). To exemplify this mechanism, we consider
the family of kernels $g_\sigma (x)=\exp[-(x/R)^\sigma]$, being
$\sigma\ge 0$ and $R$ the typical competition range. It is known
\cite{giraud} that this family of functions has non-negative
Fourier transform for $0 \le \sigma \le 2$. Interestingly enough,
the Gaussian kernel, which is the commonly adopted one
\cite{macarthurlevins,schefferpnas}, corresponds to the marginal
case. This may imply that some results previously obtained for
this case could be non robust and largely affected by the way
immigration is introduced, the presence or absence of diffusion
processes in niche space, etc.

\begin{figure}[tbh]
\includegraphics[width=8cm,clip=true]{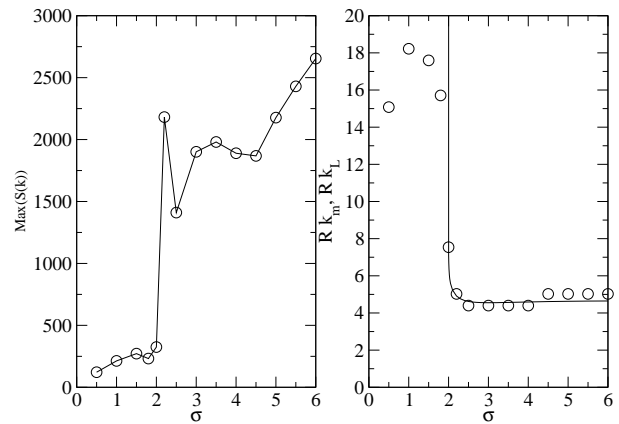}
\caption{Left panel: maximum of $\left<S(k)\right>$, the structure function
averaged over 1000 realizations of stationary distributions of species
(obtained after a time $t=10^5$, without immigration during the last half of
it) as a function of $\sigma$, for $R=0.1$, $a=0.1$. Right panel: position of
the peak $k_m$ vs $\sigma$ (circles), together with the linearly fastest
growing mode $k_L$ (line), from (\ref{dispersion}). For $\sigma>2$, the
difference between $k_m$ and $k_L$ is always smaller than the finite-size
discretization of the values of $k_m$.}
\label{fig:sf}
\end{figure}

We quantify the pattern-forming transition in terms of the
structure function, $S(k)=|\sum_j n_j \exp(i k x_j)|^2$, of the
stationary distribution of species obtained from the simulations.
The position and height of its maximum identify periodic
structures. In Fig.~\ref{fig:sf} we plot (left panel) the maximum
height of $S$ as a function of the exponent $\sigma$ of the
kernel. The sharp increase of $\max S$ for $\sigma > 2$ indicates
the formation of periodic structures in this range. This is
confirmed by the right panel plot, where we show  the position
$k_m$ of the peak of $S$, together with the value $k_L$ at which
the linear growth, expression (\ref{dispersion}), has a maximum.
Note that the location of this maximum is independent of the
parameters $a$ and $s$, being only dependent on the parameters in
$g_\sigma(x)$ ($R$ and $\sigma$; the dependence on $R$ disappears when
considering $k_L R$). The striking agreement between $k_m$ and
$k_L$ for $\sigma > 2$ confirms that the linear pattern forming
instability of the homogeneous distribution is the mechanism
responsible for the periodic species arrangement observed in that
range. Except when $\sigma\approx 2$, the value of $k_LR$ is in
the range 4.5-5.0, so that the pattern periodicity would be
$d\approx 2\pi/k_L \approx \alpha R$, with $\alpha\approx
1.3-1.4$, as observed in Fig.~\ref{fig:patterns} (right panels).

Another difference between $\sigma \le2$ and $\sigma>2$, visible in
Fig. \ref{fig:patterns}, is the existence in the later case of {\sl exclusion
zones} around established species, in which immigrants have not been able to
settle. We can understand the presence of these regions also from the density
equation (\ref{fieldtheory}), for $s=0$, by noticing that its steady stable
solutions $\phi_{st}(x)$ necessarily have regions with $\phi_{st}(x)=0$ in the
pattern forming case. This can be seen from the steady state condition $\int
dy g(|x-y|)\phi_{st}(y)=1/a$, which is valid at all niche locations $x$ at
which $\phi_{st}(x) \neq 0$. If these locations cover in fact the full niche
space $[0,1]$, we can solve the steady state condition by Fourier transform
and find that the only solution (for nonconstant $g(x)$) is the homogeneous
one $\phi_{st}(x) =(a{\mathcal N})^{-1}$. Since we know that this is linearly
unstable when the Fourier transform of $g(x)$ is not positive definite, we
conclude that steady stable solutions of (\ref{fieldtheory}) in the pattern
forming case must have regions of zero density, which we identify with the
{\sl exclusion zones}.  Given the absorbing character of the $\phi=0$ state,
many steady solutions exist, differing in the amount and location of the
$\phi_{st}=0$ segments, but the most relevant are the ones attained when $s\to
0^+$. Figure \ref{fig:patterns} (bottom right) shows one of these solutions,
numerically obtained (for a kernel $g_4(x)+\delta(x)$). The steady solution
corresponding to the $g_4(x)$ kernel of the top right panel is zero everywhere
except at a set of periodically spaced delta functions. In both cases the
discrete species distribution is well represented by the solutions of
(\ref{fieldtheory}).

\begin{figure}[tbh]
\includegraphics[width=8cm]{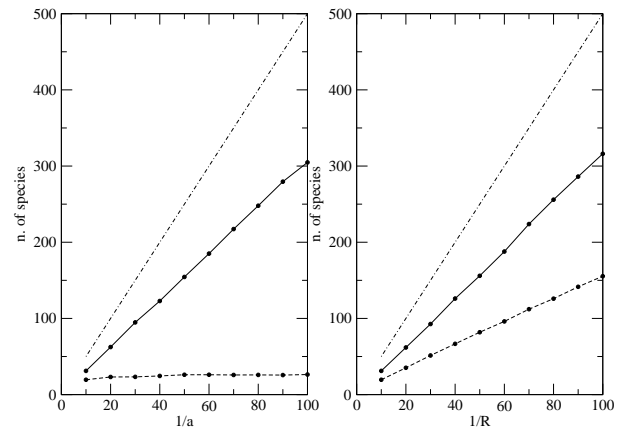}
\caption{\label{fig:NvsRa} Number of species as a function of competition
coefficient $a$ (left panel, $R=0.1$) and the interaction distance $R$ (right
panel, $a=0.1$). Symbols joined by dashed and solid lines are for the cases of
$a_i's$ heterogeneity (for which $\bar a$ is plotted instead of $a$), and
non-heterogeneity, respectively. The upper dot-dashed lines are from the
approximation (\ref{NRa}).}
\end{figure}

When $\tilde g(k)$ remains positive, as for $g_\sigma(x)$ with $\sigma \le 2$,
$\lambda(k)$ remains negative, and there are no patterns nor {\sl exclusion
zones} surviving in steady solutions of the density equation for $s\to
0^+$. Thus, the characteristic distance between species, observed in
Figs.~\ref{fig:patterns} and \ref{fig:sf} to be qualitatively different from
the case $\sigma > 2$, should be determined by a different mechanism. We
explore it for the exponential kernel, $g_1(x)=\exp(-x/R)$, because it allows
some analytical estimates. Fig. \ref{fig:NvsRa} shows the number of species at
equilibrium for $a_i=a$, and also in the heterogeneous situation in which the
$a_i$'s are independent random variables uniformly distributed between
$0.95\bar{a}$ and $1.05\bar{a}$, being $\bar{a}$ an average value.

During the system evolution (in the non-heterogeneous case) we observe that,
when the species are far from the extinction threshold, it is always possible
for a new species to successfully settle between two of them. But as a
consequence the populations of these neighboring species get reduced. This
brings the species populations closer to $n_T$ as the number of species
increases, and eventually no new species will be admitted. Thus, in this
$\sigma<2$ case, the mechanism fixing a maximum number of species, and thus a
characteristic mean distance $d$ among them, is the presence of the extinction
threshold $n_T$.

Fig.~\ref{fig:NvsRa} shows that the number of species $N$ (in the steady state
obtained after switching off immigration) grows linearly with $1/R$ and with
$1/a$ in the case of equal species, while heterogeneity slows down the
increase with $1/R$ and almost stops it with $1/\bar{a}$. Notice that
decreasing $a$ is the same as decreasing the threshold value $n_T$ due to our
rescaling of the equations. We can explain these dependences by considering an
ideal steady state made of equidistant species, at distance $d$, and having
the same population $n^*$. The equilibrium condition for the system of
equations (\ref{LVmodel}) in the exponential kernel case becomes
$\tanh(d/2R)=an^*$, which gives $d\approx 2aRn^*$ in the limit $(d/2R)\ll
1$. Recalling that $N=1/d$, each population $n^*$ decreases as the number of
species increases during immigration. The limit, setting the steady state,
will be the situation in which $n^*=n_T=1$, for which no new immigrant can be
accepted. Thus we estimate the equilibrium number of species in this case as
\begin{equation}
N\approx \left( 2aR \right)^{-1}.
\label{NRa}
\end{equation}

This is only a rough approximation, since species are not equidistant nor
equipopulated in the true equilibrium, but provides an explanation for the
observed linear scaling of $N$ with $1/a$ and $1/R$. Fig. \ref{fig:NvsRa}
shows that it gives an upper estimation for the number of species in the less
ordered distributions actually found.

The case with heterogeneity of Fig. \ref{fig:NvsRa} shows a clearly different
mechanism: the number of species does not change with $\bar{a}$ and
consequently with $n_T$. Thus, this scenario is qualitatively similar to the
pattern forming case: there is a distance in the niche space, not related to
the threshold value, of the order of the interaction range. Two species cannot
survive due to heterogeneity if they are closer than this distance,
independently on the mean competition strength.

To clarify this mechanism, we consider the role of heterogeneity in the long-range
case of a constant kernel, $g(x)=1$ for all $x$. This may be interpreted as a
case in which the kernel decaying distance goes to infinity.
Summing all the equations in (\ref{LVmodel}) we obtain an equation for the
total population $N_{tot}=\sum_j n_j$:
\begin{equation}
\dot{N}_{tot}=N_{tot}(1-\langle a \rangle N_{tot})
\end{equation}
where $\langle a \rangle=(\sum_j a_j n_j)/N_{tot}$. After a short time, the
equilibrium value $N_{tot}=\langle a \rangle^{-1}$ would be attained and we
can plug this value back into Eqs.  (\ref{LVmodel}) to obtain
$\dot{n}_i=n_i(1-a_i/\langle a \rangle)$, valid at longer times. In the case
of equal species, one has $a_i=\bar{a}=\langle a \rangle$ and all possible
states with $a\sum_j n_j=1$ are allowed. In the heterogeneous case, species
having $a_i<\langle a \rangle$ will grow while the others will decrease their
population and finally go extinct. Meanwhile, it is easy to realize that
$\langle a \rangle$ will increase, sending more and more species below the
extinction threshold. The final result, valid for any initial distribution of
the $a_i$'s, is that just one species will survive, as confirmed by
simulations (not shown).

%

To conclude, we studied analytically and numerically the collective behavior
of competitive Lotka-Volterra systems. Our main message is that the form of
the competition kernel changes drastically the equilibrium distribution of
species. Species clustering with periodic spacings of the order of the
interaction range can occur at one side of a pattern forming transition,
whereas smaller spacings, depending on the interaction strength $a$, occur at
the other. Surprisingly, the Gaussian kernel, the one usually considered in
the literature, corresponds to a frontier case. Diversity has been shown to
alter qualitatively the competition outcome. Generalization to the case of a
multidimensional niche space is straightforward and does not change
qualitatively our results. Diffusion in niche space, modelling mutations
\cite{schefferpnas}, can be introduced and has a stabilizing effect somehow
similar to that of the immigration rate $s$.

\acknowledgments
 Financial support from FEDER and MEC (Spain),
through project CONOCE2 (FIS2004-00953) is greatly acknowledged.


\begin{thebibliography}{200}

\bibitem{macarthurlevins} R. MacArthur and R. Levins, Am. Nat. {\bf 101}, 377 (1967).

\bibitem{hardin} G. Hardin, Science {\bf 131}, 1292 (1960).

\bibitem{szabo} P. Szab\'o and G. Mesz\'ena, Oikos {\bf 112}, 612 (2006).

\bibitem{gyllenberg}M. Gyllenberg and G. Mesz\'ena, J. Math. Biol. {\bf
  50}, 133 (2005).

\bibitem{schefferpnas} M. Scheffer and E.H. Van Nes, Proc. Natl. Ac. Sci. {\bf
  103}(16), 6230 (2006).

\bibitem{brigatti} E. Brigatti, J.S.S. Martins and I. Roditi, Physica A, in press.

\bibitem{johansson} J. Johansson and J. Ripa, Am. Nat. {\bf 168}, 572 (2006).

\bibitem{tokitaprl} K. Tokita, Phys. Rev. Lett. {\bf 93}, 178102 (2004).

\bibitem{optics} C. Benkert and D.Z. Anderson, Phys. Rev. A {\bf
44}, 4633 (1991).

\bibitem{tech} C.W.I. Pistorius and J.M. Utterback, Res. Policy {\bf
26}, 67 (1997).

\bibitem{Kurtze} D.A. Kurtze, Phys. Rev. B {\bf 40}, 11104 (1989).

\bibitem{refbugs1} E. Hern\'andez-Garc\'\i a and C. L\'opez, Phys. Rev. E {\bf
70} 016216 (2004);
C. L\'opez and E. Hern\'andez-Garc\'\i a, Physica D {\bf
199}, 223 (2004).

\bibitem{macarthurwilson} R. MacArthur and E. Wilson, {\em The theory of
island biogeography}, Princeton University Press (1967);
S.P. Hubbell {\em  The Unified Neutral Theory of Biodiversity
  and Biogeography}, Princeton University Press (2001);
I. Volkov, J.R. Banavar, S.P. Hubbell, A. Maritan, Nature
  {\bf 424}(6952), 1035 (2003);
S. Pigolotti, A. Flammini, A. Maritan, Phys. Rev. E {\bf 70}, 011916 (2004).

\bibitem{CrossHoh} M.C. Cross and P.C. Hohenberg, Rev. Mod. Phys. {\bf 65}, 851
(1993).

\bibitem{giraud} B.G. Giraud, Acta Phys. Polon. B {\bf 37}, 331 (2006).

\end{thebibliography}
\end{document}